\documentclass[12pt]{article}

\usepackage{graphicx}
\usepackage{amsmath}
\usepackage{amssymb}
\usepackage{amscd}
\usepackage{latexsym}

\usepackage{cite}

\newcommand{\be}{\begin{equation}}
\newcommand{\ee}{\end{equation}}
\newcommand{\br}{{\bf r}}
\newcommand{\bk}{{\bf k}}

\newcommand{\sgm}{\sigma}
\newcommand{\vp}{\varphi}
\newcommand{\om}{\omega}
\newcommand{\ep}{\varepsilon}
\newcommand{\lbd}{\lambda}
\newcommand{\Lbd}{\Lambda}

\newcommand{\gm}{\gamma}

\newcommand{\dlt}{\delta}
\newcommand{\cH}{{\cal H}}
\newcommand{\cF}{{\cal F}}
\newcommand{\rgl}{\rangle}
\newcommand{\lgl}{\langle}
\newcommand{\ra}{\rightarrow}
\newcommand{\dgr}{\dagger}
\newcommand{\prt}{\partial}

\begin{document}

\begin{center}
{\Large{\bf 
Mid-range order in trapped quasi-condensates of bosonic atoms} \\ [5mm]

V.I. Yukalov$^{1,2}$ and E.P. Yukalova$^3$} \\ [3mm]

{\it
$^1$Bogolubov Laboratory of Theoretical Physics, \\
Joint Institute for Nuclear Research, Dubna 141980, Russia \\ [2mm]
                                      
$^2$Instituto de Fisica de S\~{a}o Carlos, Universidade de S\~{a}o Paulo, CP 369,  \\
S\~{a}o Carlos 13560-970, S\~{a}o Paulo, Brazil  \\ [2mm]

$^3$Laboratory of Information Technologies, \\
Joint Institute for Nuclear Research, Dubna 141980, Russia} \\ [5mm]

{\bf E-mails}: yukalov@theor.jinr.ru, yukalova@theor.jinr.ru 

\end{center}


\vskip 2cm

\begin{abstract}
Finite Bose systems cannot display a genuine Bose-Einstein condensate with infinite long-range
order. But, if the number of trapped atoms is sufficiently large, a kind of Bose-Einstein 
condensation does occur, with the properties of the arising quasi-condensate being very close 
to the genuine condensate. Although the quasi-condensate does not enjoy long-range order, it
has mid-range order. This paper shows that the level of mid-range order in finite Bose systems 
can be characterized by order indices of density matrices.  
\end{abstract}

\section{Introduction}

Bose-Einstein condensation of trapped atoms has been a hot topic in recent years, intensively
investigated both experimentally and theoretically (see, e.g., review articles and books 
\cite{Courteille_1,Andersen_2,Yukalov_3,Bongs_4,Yukalov_5,Lieb_6,Posazhennikova_7,Morsch_8,
Yukalov_9,Letokhov_10,Moseley_11,Bloch_12,Proukakis_13,Yurovsky_14,Pethick_15,Yukalov_16,
Yukalov_17,Yukalov_18,Yukalov_19}). Trapped atoms represent finite systems. In finite systems,
strictly speaking, there can be no phase transitions with arising long-range order. Respectively,
there can be no genuine phase transition of Bose-Einstein condensation in a finite trap. 
Nevertheless, when a finite system is sufficiently large, its properties can be so close to a
bulk system that, with a good approximation, one can speak of phase transitions there. In that
sense, one studies Bose-Einstein condensation in finite traps. And one calls the arising 
quasi-condensate just Bose condensate. 

However, it is interesting as to what extent the quasi-condensate in a finite system differs 
from the genuine condensate. More precisely, how it would be possible to distinguish infinite 
off-diagonal order, corresponding to the genuine condensate accompanied by spontaneous breaking 
of global gauge symmetry, from quasi-condensate in a finite system? Is it possible to define a
measure quantifying the level of a quasi-long-range order?

In the present paper, we show that the Bose quasi-condensate in a finite system is characterized
by mid-range order that can be quantified by order indices.

\section{Order indices}

Order indices were introduced for density matrices in \cite{Coleman_20} and considered for 
several macroscopic systems \cite{Coleman_21,Coleman_22,Coleman_23,Coleman_24}. The notion of 
order indices was generalized for arbitrary matrices and operators in \cite{Yukalov_25}.     
 
Let an operator $\hat{A}$ acting on a Hilbert space $\mathcal{H}$ possess a finite norm 
$\|\hat{A}\|$ and a trace. The order index of the operator is defined as
\be
\label{1}
 \om(\hat A) \equiv \frac{\log||\;\hat A \;||}{\log|\; {\rm Tr}_\cH \hat A \;| } \;  .
\ee
One says that an operator $\hat{A_1}$ is better ordered than $\hat{A_2}$ if and only if the
order index of $\hat{A_1}$ is larger than that of $\hat{A_2}$. If their order indices are 
equal, then one says that these operators are equally ordered. The convenient norm in 
definition (\ref{1}) is the Hermitian operator norm
$$
||\;\hat A \;|| = \sup_\vp \; \frac{|\; \hat A\vp\;| }{|\;\vp\;| } =
\sup_\vp \left[ \frac{ (\hat A\vp,\hat A\vp) }{ (\vp,\vp) } \right]^{1/2} \;  ,
$$
where $\vp\in\mathcal{H}$ is not zero. If the operator is Hermitian and $\{\vp_k\}$ is 
an orthonormal basis in $\mathcal{H}$, then the Hermitian norm reduces to
$$
||\;\hat A \;|| = \sup_\vp \; \frac{|(\vp,\hat A\vp)| }{|\;\vp\;|} = 
\sup_{\vp_k}(\vp_k,\hat A\vp_k) \qquad ( \hat A^+ = \hat A ) \; .
$$
For a semi-positive operator $\hat{A} \geq 0$ the norm is not larger than the trace, because
of which
\be
\label{2}
 \om(\hat A) \leq 1 \qquad ( \hat A \geq 0 ) \;  .
\ee

Order indices can be introduced for generalized density matrices as follows. Let $x$ be a set
of physical variables and $\mathcal{A} = \{\hat{A}(x)\}$ be the algebra of local observables
acting on a Fock space $\mathcal{F}$. And let 
$\mathcal{A}_\psi = \{\hat{\psi}(x),\hat{\psi}^\dagger(x)\}$ be the algebra of field operators
on $\mathcal{F}$. The union $\mathcal{A}_{ext} \equiv \mathcal{A} \bigoplus \mathcal{A}_\psi$
is called the extended local algebra. 

For any representative $A(x)$ of the extended local algebra $\mathcal{A}_{ext}$, it is possible 
to define the averages
$$
D_A(x_1,x_2,\ldots,x_n;y_1,y_2,\ldots,y_n) \; =
$$
\be
\label{3}
 = \;
{\rm Tr}_\cF A(x_1)A(x_2)\ldots A(x_n) \; \hat\rho \;
A^+(y_n)A^+(y_{n-1})\ldots A^+(y_1) \;  ,
\ee
where $\hat{\rho}$ is a statistical operator. These averages can be treated as matrix elements 
of the matrix
\be
\label{4}
 \hat D_A^n \equiv [ \; D_A(x_1,x_2,\ldots,x_n;y_1,y_2,\ldots,y_n)\; ] \;  ,
\ee
which can be called a generalized density matrix.   

It is also possible to consider the functions $\varphi(x_1,x_2,\ldots,x_n)$ of a Hilbert space
$\mathcal{H}_n$ as columns
\be
\label{5}
  \vp_n \equiv [ \; \vp(x_1,x_2,\ldots,x_n)\; ] \; .
\ee
Then the norm of matrix (\ref{4}) can be defined as
\be
\label{6}
 || \; \hat D_A^n\;|| = \sup_{\vp_n} \; \frac{(\vp_n,\hat D_A^n\vp_n)}{(\vp_n,\vp_n)} \; ,
\ee
where the scalar product is given by the definition
$$
 (\vp_n,\vp_n') = \int \vp^*(x_1,x_2,\ldots,x_n) \vp_n'(x_1,x_2,\ldots,x_n) \;
\prod_{i=1}^n dx_i \; .
$$
Respectively, the trace of the matrix is
\be
\label{7}
 {\rm Tr} \hat D_A^n = \int D_A (x_1,x_2,\ldots,x_n;x_1,x_2,\ldots,x_n)\; 
\prod_{i=1}^n dx_i \; .
\ee
The order index of the generalized density matrix is
\be
\label{8}
 \om(\hat D_A^n) \equiv \frac{\log||\;\hat D_A^n\;||}{\log|\; {\rm Tr}\hat D_A^n\;|} \; .
\ee

A particular case of the generalized density matrices are the reduced density matrices
\cite{Coleman_26}
\be
\label{9}
 \hat\rho_n = \hat D_\psi^n = [\;\rho(x_1,x_2,\ldots,x_n;y_1,y_2,\ldots,y_n)\;] \; ,
\ee
with the matrix elements
$$
 \rho(x_1,x_2,\ldots,x_n;y_1,y_2,\ldots,y_n) \; \equiv
$$
\be
\label{10}
\equiv \;
 {\rm Tr}_\cF \hat\psi(x_1) \hat\psi(x_2)\ldots \hat\psi(x_n) \; \hat\rho \; 
\hat\psi^\dgr(y_n) \hat\psi^\dgr(y_{n-1}) \ldots \hat\psi^\dgr(y_1) \; .
\ee
The related order index of a density matrix (\ref{9}) is 
\be
\label{11}
 \om(\hat\rho_n) \equiv \frac{\log||\;\hat\rho_n\;||}{\log|\; {\rm Tr}\hat\rho_n \;|} \;  .
\ee

If $\vp_{nk}$ is an eigenfunction, labeled by a multi-index $k$, of the density matrix (\ref{9}), 
then its eigenvalues are 
\be
\label{12}
  N_{nk} \equiv ( \vp_{nk} , \hat\rho_n\vp_{nk} ) 
\ee
and its Hermitian norm is
\be
\label{13}
 ||\;\hat\rho_n\;|| = \sup_k N_{nk} \;  .
\ee
Because of the trace
$$
 {\rm Tr} \hat\rho_n = \frac{N!}{(N-n)!} \;  ,   
$$
the order index (\ref{11}) can be represented in the form
\be
\label{14}
 \om(\hat\rho_n) = \frac{\log\sup_k N_{nk}}{\log[N!/(N-n)!]} \;  .
\ee

The norms of the reduced density matrices satisfy \cite{Coleman_26} the inequalities
$$
||\; \hat\rho_n \; || \; \leq \; ( b_n N )^n \qquad (Bose)
$$
for Bose particles and
$$
||\; \hat\rho_{2n} \; || \; \leq \; ( c_{2n} N )^n \; , \qquad 
||\; \hat\rho_{2n+1} \; || \; \leq \; ( c_{2n+1} N )^n \qquad (Fermi) \; ,
$$
for Fermi particles, where $b_n$ and $c_n$ are finite numbers. Therefore for the order indices 
of density matrices, under large $N \gg 1$,
we have
\be
\label{15}
 \om(\hat\rho_n) \; \leq \; 1 \qquad (Bose)
\ee
for bosons and
\be
\label{16}
\om(\hat\rho_{2n}) \; \leq \; \frac{1}{2} \; , \qquad 
\om(\hat\rho_{2n+1}) \; \leq \; \frac{n}{2n+1} \qquad (Fermi)
\ee
for fermions.

\section{Dilute gas}

As an example of a concrete finite system, let us consider a dilute Bose gas with local 
interactions
\be
\label{17}
 \Phi(\br) = \Phi_0\dlt(\br) \; , \qquad \Phi_0 \equiv 4\pi \; \frac{a_s}{m} \;  ,
\ee
in which $m$ is atomic mass and $a_s$, scattering length. Here and in what follows, the Planck 
and Boltzmann constants are set to one, $\hbar = 1$ and $k_B = 1$. The energy Hamiltonian reads 
as 
\be
\label{18} 
\hat H = \int \hat\psi^\dgr(\br) 
\left( -\; \frac{\nabla^2}{2m} \right) \psi(\br) \; d\br  \; +  \; 
\frac{1}{2} \; \Phi_0 
\int \hat\psi^\dgr(\br) \hat\psi^\dgr(\br) \hat\psi(\br) \hat\psi(\br)\; d\br \; ,
\ee
where $\hat{\psi}({\bf r})$ is a Bose field operator, generally depending on time $t$, which
is not shown for simplicity of notation. 

Below we employ the self-consistent approach, reviewed in 
\cite{Yukalov_16,Yukalov_17,Yukalov_18,Yukalov_27}, guaranteeing the correct description 
of Bose-condensed systems. This approach possesses several unique features: (i) satisfies 
all conservation laws; (ii) gives a gapless spectrum; (iii) describes Bose-Einstein condensation 
as a second order phase transition; (iv) provides for the condensate fraction, as a function 
of interaction strength, good numerical agreement with Monte Carlo simulations, both for uniform 
as well as for trapped systems; (v) leads to the behavior of the ground state energy, under 
varying interaction strength, which, at weak interactions, yields exactly the Lee-Huang-Yang 
formula \cite{Lee_28,Lee_29,Lee_30} and, at strong interactions, it is close to the results of 
Monte Carlo calculations; (vi) explains the effect of local condensate depletion at trap center 
under strong interactions \cite{Yukalov_31}, agreeing well with Monte Carlo simulations.   

We start with the Bogolubov shift \cite{Bogolubov_32,Bogolubov_33,Bogolubov_34}
\be
\label{19}
\hat\psi(\br) = \eta(\br) + \psi_1(\br)
\ee
separating the condensate (quasi-condensate) function $\eta$ from the field operator of 
uncondensed atoms $\psi_1$, with these quantities being mutually orthogonal,
\be
\label{20}
 \int \eta^*(\br) \psi_1(\br) \; d\br = 0 \; .
\ee
The condensate function plays the role of an order parameter, so that
\be
\label{21}
 \eta(\br) = \lgl \; \hat\psi(\br)\; \rgl \; , \qquad 
\lgl \; \psi_1(\br)\; \rgl = 0 \; .
\ee
The total number of atoms $N = N_0 + N_1$ is formed by the number of condensed atoms
\be
\label{22}
N_0 = \int |\;\eta(\br)\;|^2 \; d\br
\ee
and the number of uncondensed atoms
\be
\label{23}
 N_1 = \lgl \; \hat N_1\; \rgl \; , \qquad 
\hat N_1 \equiv \int \psi_1^\dgr(\br) \psi_1(\br) \; d\br \; .
\ee

The grand Hamiltonian, taking into account conditions (\ref{21}), (\ref{22}), and 
(\ref{23}), reads as
\be
\label{24}
 H = \hat H - \mu_0 N_0 - \mu_1 \hat N_1 - \hat\Lbd \;  ,
\ee
where
$$
 \hat\Lbd = \int [\; \lbd(\br) \psi_1^\dgr(\br) + \lbd^*(\br) \psi_1(\br)\; ] \; d\br \; .
$$
The quantities $\mu_0$, $\mu_1$, and $\lambda({\bf r})$ are the Lagrange multipliers 
guaranteeing the validity of these conditions. 

Equations of motion, equivalent to the Heisenberg equations \cite{Yukalov_17}, are the 
equation for the condensate fraction
\be
\label{25}
i\; \frac{\prt}{\prt t}\; \eta(\br,t) = 
\left\lgl \frac{\dlt H}{\dlt\eta^*(\br,t)}\right\rgl
\ee
and the equation for the field operator of uncondensed atoms
\be
\label{26}
i\; \frac{\prt}{\prt t}\; \psi_1(\br,t) =  \frac{\dlt H}{\dlt\psi_1^\dgr(\br,t)} \; .
\ee

The first-order density matrix is
\be
\label{27}
  \hat\rho_1 = [\; \rho(\br,\br')\; ] \; , \qquad 
\rho(\br,\br') = \lgl \; \hat\psi^\dgr(\br')\hat\psi(\br) \; \rgl \; .
\ee
With the Bogolubov shift (\ref{19}), we have
\be
\label{28}
 \rho(\br,\br') = \eta^*(\br')\eta(\br) + 
\lgl \; \hat\psi_1^\dgr(\br')\hat\psi_1(\br) \; \rgl \;  .
\ee
The eigenvalues of the density matrix are
\be
\label{29}
  N_{1k} = \int \vp_k^*(\br) \rho(\br,\br') \vp_k(\br')\; d\br d\br' \; ,
\ee
provided that $\varphi_k$ are the eigenfunctions. Using (\ref{28}), the eigenvalues can 
be written as the sum
\be
\label{30}
N_{1k} = N_k + n_k \;   ,
\ee
in which 
\be
\label{31}
N_k \equiv \left| \int \eta^*(\br)\vp_k(\br)\; d\br \right|^2
\ee
and
\be
\label{32}
 n_k = \lgl a_k^\dgr a_k \rgl \; , \qquad 
a_k \equiv \int \vp_k^*(\br) \psi_1(\br) \; d\br \;  .
\ee

In what follows, we shall study the order index of the single-particle density matrix
\be
\label{33}
 \om(\hat\rho_1) = \frac{\log||\;\hat\rho_1\;||}{\log N} \;  ,
\ee
with the norm of matrix (\ref{27}) 
$$
||\;\hat\rho_1\;|| = \sup_k N_{1k} = \sup_k (N_k + n_k) \;   .
$$

\section{Box-shaped trap}

We consider a trap having the shape of a box of volume $V = L^3 = N a^3$. Bose-Einstein 
condensate (quasi-condensate) in such a box trap has recently been observed \cite{Navon_35}. 
As usual, the box is assumed to be periodically continued. 

The eigenfunctions of the first-order density matrix are plane waves
\be
\label{34}
 \vp_k(\br) = \frac{1}{\sqrt{V}} \; e^{i\bk\cdot\br} \;  .
\ee
The condensate function becomes a constant
\be
\label{35}
 \eta(\br) = \sqrt{\frac{N_0}{V} } \;  .
\ee
The density-matrix eigenvalues are
\be
\label{36}
 N_{1k} = N_0 \dlt_{k0} + n_k \;  ,
\ee
which defines the norm 
\be
\label{37}
||\;\hat\rho_1\;|| = \sup\{N_0, \; \sup_k n_k \} \;   .
\ee

The average atomic density 
\be
\label{38}
\rho \equiv \frac{N}{V} = \rho_0 + \rho_1
\ee
is the sum of the condensate density $\rho_0$ and the density of uncondensed atoms $\rho_1$,
\be
\label{39}
 \rho_0 \equiv \frac{N_0}{V} \; , \qquad   
\rho_1 \equiv \frac{N_1}{V} = \frac{1}{V} \sum_k n_k \; .
\ee

Employing the Hartree-Fock-Bogolubov decoupling, we find \cite{Yukalov_16,Yukalov_17,Yukalov_18}
the distribution of uncondensed atoms
\be
\label{40}
 n_k = \frac{\om_k}{2\ep_k} \; \coth\left( \frac{\ep_k}{2T}\right) \; - \; \frac{1}{2} \;  ,   
\ee
where we use the notation
\be
\label{41}
 \om_k \equiv mc^2 + \frac{k^2}{2m} \;  ,
\ee
$\varepsilon_k$ is the spectrum of collective excitations
\be
\label{42}
\ep_k = \sqrt{(ck)^2 + \left(\frac{k^2}{2m}\right)^2 } \;   ,
\ee
and the sound velocity $c$ satisfies the equation
\be
\label{43}
  mc^2 = \Phi_0 ( \rho_0 + \sgm_1) \; .
\ee
Here the anomalous average is 
\be
\label{44}
 \sgm_1 = \frac{1}{V} \sum_k \sgm_k \; ,  \qquad 
\sgm_k = - \; \frac{mc^2}{2\ep_k} \; \coth\left( \frac{\ep_k}{2T}\right) \; .
\ee

Considering zero temperature, we get 
\be
\label{45}
 n_k = \frac{\om_k -\ep_k}{2\ep_k} \; , \qquad \sgm_k = -\; \frac{mc^2}{2\ep_k} 
\qquad ( T = 0 ) \;  ,
\ee
the density of uncondensed atoms
\be
\label{46}
 \rho_1 = \frac{(mc)^3}{3\pi^2} \qquad ( T = 0 ) \;  ,
\ee
and the anomalous average
\be
\label{47}
 \sgm_1 = - mc^2 \int \frac{1}{2\ep_k} \; \frac{d\bk}{(2\pi)^3} \qquad ( T = 0 ) \;  .
\ee

The anomalous average (\ref{47}) diverges and requires a regularization. This can be done by
resorting to the dimensional regularization, that provides asymptotically exact results at low
density and weak interactions, and then accomplishing analytic continuation extending the 
results to finite density and interaction strength \cite{Yukalov_36}. 

We notice that at asymptotically weak interaction, equation (\ref{43}) leads to the Bogolubov
sound velocity
\be
\label{48}
 c \ra c_B \equiv \sqrt{\frac{\rho}{m}\; \Phi_0} \qquad ( \rho \Phi_0 \ra 0 ) \;  .
\ee
Because of this, at small values of $\rho \Phi_0$, the anomalous average (\ref{44}) can be
represented as
\be
\label{49}
 \sgm_1 \simeq - mc_B^2 \int \frac{1}{2\ep_k} \; \frac{d\bk}{(2\pi)^3} \qquad 
( \rho\Phi_0 \ra 0 ) \;  .
\ee
Using the dimensional regularization \cite{Andersen_2,Yukalov_18,Kleinert_37} yields
\be
\label{50}
 \int \frac{1}{2\ep_k} \; \frac{d\bk}{(2\pi)^3} = - \; \frac{m^2c^{(n)}}{\pi^2} \qquad
( \rho\Phi_0 \ra 0 ) \;   ,
\ee
where $c^{(n)}$ is a weak-interaction approximation of sound velocity. The latter, together
with (\ref{49}), gives the corresponding approximation for the anomalous average
\be
\label{51}
\sgm_1^{(n)} = \frac{m^3c_B^2}{\pi^2} \; c^{(n)} \;   .
\ee
To analytically continue the sound velocity $c^{(n)}$ to finite values of $\rho\Phi_0$,
we employ equation (\ref{43}) in the form
\be
\label{52}
c^{(n+1)}   = 
\left[\; \frac{\Phi_0}{m} \left( \rho_0 + \sgm_1^{(n)} \right) \; \right]^{1/2} \; .
\ee
Combining (\ref{51}) and (\ref{52}) gives the iterative equation
\be
\label{53}
 \sgm_1^{(n+1)} = \frac{(mc_B)^3}{\pi^2\sqrt{\rho}} \; 
\sqrt{\rho_0 + \sgm_1^{(n)} } \; .
\ee
Defining dimensionless fractions
\be
\label{54}
n_0 \equiv \frac{\rho_0}{\rho} \; , \qquad n_1 \equiv \frac{\rho_1}{\rho} \; , 
\qquad \sgm \equiv \frac{\sgm_1}{\rho}
\ee
and dimensionless sound velocities
\be
\label{55}
s \equiv \frac{mc}{\rho^{1/3}} \; , \qquad s_B \equiv \frac{mc_B}{\rho^{1/3}}
\ee
reduces (\ref{53}) to the dimensionless equation
\be 
\label{56}
\sgm^{(n+1)} =  \frac{s_B^3}{\pi^2} \; \sqrt{n_0 + \sgm^{(n)} } \; .
\ee
It is reasonable to start the iterative procedure from the Bogolubov approximation, that 
is asymptotically exact at low $\rho \Phi_0 \ra 0$, where $\sigma^{(0)} = 0$. In the second 
order, we obtain
\be
\label{57}
\sgm = \frac{s_B^3}{\pi^2} \; 
\left( n_0 + \frac{s_B^3}{\pi^2}\; \sqrt{n_0} \right)^{1/2} \;   .
\ee
This anomalous average will be used below.

\section{Order-index behaviour}

Now we shall calculate the order index (\ref{33}) for a finite box, where we take the natural
logarithms. For a periodically continued finite system there exists the minimal wave vector
\be
\label{58}
 k_{min} = \frac{2\pi}{L} =  2\pi\left( \frac{\rho}{N}\right)^{1/3} \;  .
\ee
The maximal value of $n_k$ occurs at the minimal wave vector,
$$
 \sup_k n_k = \sup_k \; \frac{\om_k-\ep_k}{2\ep_k} = \frac{mc}{2k_{min}} \;  ,
$$  
which yields
\be
\label{59}
  \sup_k n_k = \frac{s}{4\pi} \; N^{1/3} \;  .
\ee
Thus the norm of the first-order density matrix is
\be
\label{60}
 ||\;\hat\rho_1\;|| = \sup\left\{ n_0 N , \; \frac{s}{4\pi}\; N^{1/3}\right\} \;  .
\ee
In that way, we need to study the behavior of the order index
\be
\label{61}
 \om(\hat\rho_1) = \sup\left\{ 1 + \frac{\ln n_0}{\ln N} \; , \; 
\frac{1}{3} + \frac{\ln(s/4\pi)}{\ln N} \right\} \;  .
\ee

It is convenient to introduce the dimensionless gas parameter
\be
\label{62}
\gm \equiv a_s \rho^{1/3}
\ee
characterizing the interaction strength. Then the Bogolubov sound velocity takes the 
form $s_B=\sqrt{4\pi\gm}$. And equation (\ref{43}) becomes
\be
\label{63}
 s^2 = 4\pi\gm ( n_0 + \sgm) \;  .
\ee
The condensate fraction is
\be
\label{64}
n_0 =  1 \; - \; \frac{s^3}{3\pi^2} \; ,
\ee
and the anomalous average (\ref{57}) reduces to
\be
\label{65}
\sgm = \frac{8}{\sqrt{\pi}} \; \gm^{3/2} \left[ \; n_0 + 
 \frac{8}{\sqrt{\pi}} \; \gm^{3/2}\; \sqrt{n_0} \;\right]^{1/2} \;  .
\ee
 
At small gas parameter $\gamma \ra 0$, the condensate fraction can be expanded as
\be
\label{66}
 n_0 \simeq 1 \; - \; \frac{8}{3\sqrt{\pi}}\; \gm^{3/2} \; - \; 
\frac{64}{3\pi}\; \gm^3 \; - \; \frac{256}{9\pi^{3/2}}\; \gm^{9/2} \; + \;
\frac{61952}{81\pi^2}\; \gm^6 \;   .
\ee
The anomalous average has the expansion
\be
\label{67}
 \sgm \simeq  \frac{8}{\sqrt{\pi}}\; \gm^{3/2} \; + \; 
\frac{64}{3\pi}\; \gm^3 \; - \; \frac{1408}{9\pi^{3/2}}\; \gm^{9/2} \; - \;
\frac{1792}{27\pi^2}\; \gm^6 \; .
\ee
While the sound velocity behaves as
\be
\label{68}
 s \simeq \sqrt{4\pi\gm} \; + \; \frac{16}{3}\; \gm^2 \; - \; 
\frac{64}{9\sqrt{\pi}}\; \gm^{7/2} \; - \; \frac{4480}{27\pi}\; \gm^5  \; .
\ee
In numerical form, the expansions are:
$$
n_0 \simeq 1 - 1.50451 \gm^{3/2} - 6.79061\gm^3 - 5.10826\gm^{9/2} + 77.4944\gm^6 \; ,
$$
$$
\sgm \simeq 4.51352\gm^{3/2} + 6.79061\gm^3 - 28.0954 \gm^{9/2} - 6.72472 \gm^6 \; ,
$$
$$
 s \simeq 3.54491\gm^{1/2} + 5.33333 \gm^2 - 4.01201\gm^{7/2} - 52.8159\gm^5 \;  .
$$

For the order index, we find 
\be
\label{69}
  \om(\hat\rho_1) \simeq 1 \; - \; \frac{8}{3\sqrt{\pi}\ln N} \; \gm^{3/2} \; - \;
\frac{224}{9\pi\ln N} \; \gm^3 \qquad (\gm \ra 0) \; .
\ee

At strong interaction, when $\gamma \ra \infty$, the condensate fraction reads as 
\be
\label{70}
 n_0 \simeq 0.397978 \; \frac{10^{-4}}{\gm^{13}} \; - \; 
0.111251\; \frac{10^{-6}}{\gm^{21}} \;  ,
\ee
the anomalous average is
\be
\label{71}
\sgm \simeq 0.761618 \; \frac{1}{\gm} \; - \; 
0.397978\; \frac{10^{-4}}{\gm^{13}} \; - \; 
0.202072\; \frac{10^{-4}}{\gm^{14}} \; + \; 
0.111251\; \frac{10^{-6}}{\gm^{21}} \;   ,
\ee
and the sound velocity has the expansion
\be
\label{72}
 s \simeq 3.09367 \; - \; 0.410404\;  \frac{10^{-4}}{\gm^{13}} \; + \; 
0.114725 \;  \frac{10^{-6}}{\gm^{21}} \;  .
\ee

The order index behaves as 
\be 
\label{73}
 \om(\hat\rho_1) \simeq \frac{1}{3} \; - \; \frac{1.40167}{\ln N} \; - \; 
\frac{0.132659}{\ln N} \; \frac{10^{-4}}{\gm^{13}} \qquad ( \gm \ra \infty) \; .
\ee

In this way, the limits of large $N$ and large $\gamma$ are not commutative. The limit of 
large $N$, for any finite $\gamma$, gives 
\be
\label{74}
 \lim_{N\ra\infty} \; \om(\hat\rho_1) = 1 \qquad ( \gm < \infty) \;  ,
\ee
which defines the genuine Bose condensate in thermodynamic limit. While, if we first take 
the limit of large $\gamma$, under finite $N$, and after this, the limit of large $N$, we 
get
\be
\label{75}
 \lim_{N\ra\infty} \; \lim_{\gm\ra\infty} \; \om(\hat\rho_1) = \frac{1}{3} \;  .
\ee
For finite $N$ and $\gamma$, the order index is smaller than one, which implies 
that we have not a genuine condensate, with a long-range off-diagonal order, but a 
quasi-condensate possessing only mid-range order.  

The influence of varying the interaction strength $\gamma$ and the number of trapped 
atoms $N$ on the system characteristics is illustrated in Figs. 1 to 3. Figure 1 describes 
the dependence of the quasi-condensate fraction $n_0$, anomalous average $\sigma$, and 
sound velocity $s$ on the strength of the gas parameter $\gamma$. Figure 2 shows that 
increasing $\gamma$ diminishes the order index, which is quite natural, since strong 
interactions are known to deplete the condensate. And Fig. 3 demonstrates that increasing 
the number of trapped atoms leads to larger values of the order index.

In conclusion, we have shown that in finite quantum systems, where there is no long-range 
order, there can exist mid-range order, which can be quantified by order indices of density 
matrices. Finite systems of trapped Bose atoms can exhibit quasi-condensate possessing 
mid-range order, with an order index smaller than one. The order index of the first-order 
density matrix shows the relation between the norm of the matrix and the number of atoms 
in the system,
$$
||\; \hat\rho_1\; || = N^{\om(\hat\rho_1)} \;   .
$$  
For a large number of atoms in a trap, the order index can be so close to unity that the 
quasi-condensate becomes almost indistinguishable from the genuine condensate. But for a 
not so large number of atoms, or very strong interactions, the order index can essentially 
deviate from unity.

\newpage

\begin{center}
{\Large {\bf Figure Captions}}
\end{center}

\vskip 2cm
{\bf Figure 1}. Condensate (quasi-condensate) fraction $n_0$ (dash-dotted line), 
anomalous average $\sgm$ (dashed line), and sound velocity $s$ (solid line) as 
functions of the gas parameter $\gm$.

\vskip 2cm
{\bf Figure 2}. Order index $\om(\hat{\rho}_1)$ as a function of the gas parameter 
$\gm$ for different numbers of trapped atoms: $N=10$ (solid line), $N=10^3$ (dashed 
line), and $N=10^6$ (dash-dotted line).

\vskip 2cm
{\bf Figure 3}. Order index $\om(\hat{\rho}_1)$ as a function of $\ln N$ for 
different values of the gas parameter: $\gm=0.1$ (solid line), $\gm=0.5$ (dashed 
line), $\gm=1$ (dash-dotted line), and $\gamma = 2$ (dotted line).

\newpage

\begin{figure}[ht]
\centerline{\includegraphics[width=10cm]{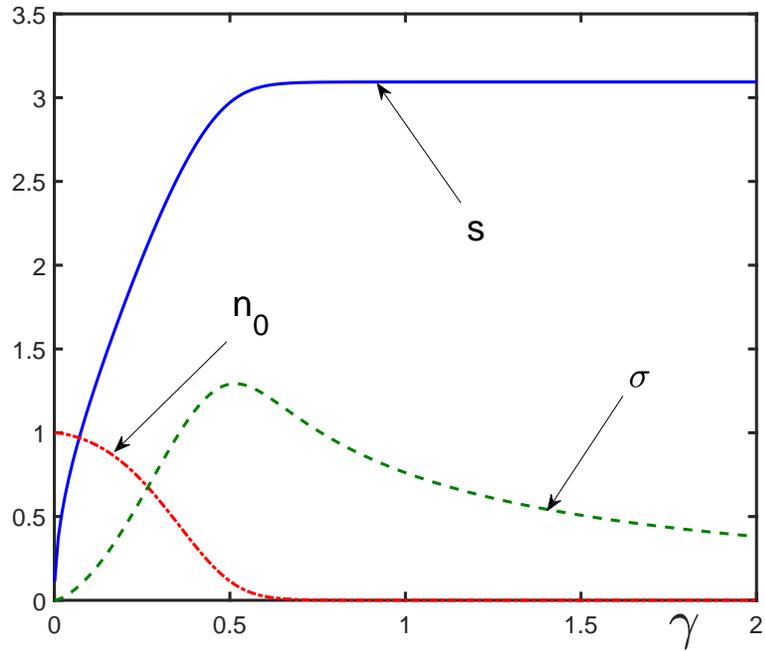}}
\caption{Condensate (quasi-condensate) fraction $n_0$ (dash-dotted line), anomalous 
average $\sgm$ (dashed line), and sound velocity $s$ (solid line) as functions of the 
gas parameter $\gm$.}
\label{fig:Fig.1}
\end{figure}

\begin{figure}[ht]
\centerline{\includegraphics[width=10cm]{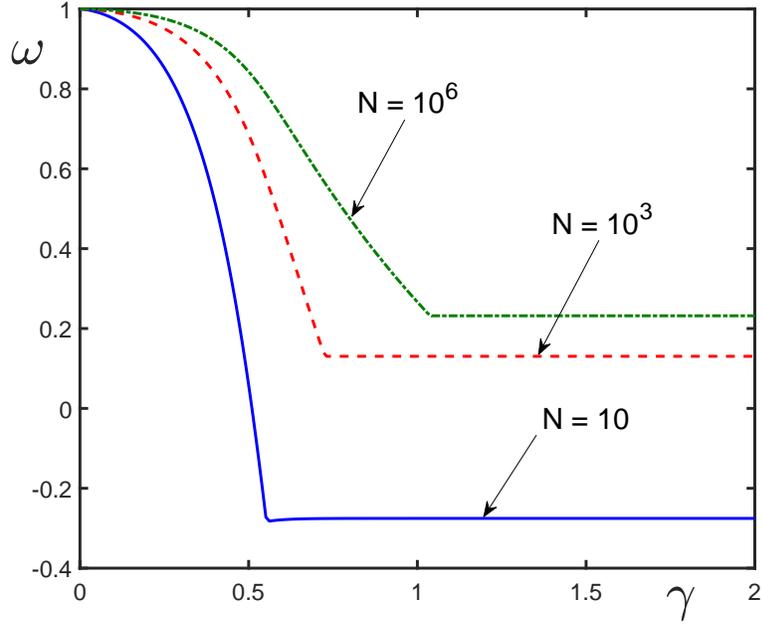}}
\caption{Order index $\om(\hat{\rho}_1)$ as a function of the gas parameter $\gm$ for
different numbers of trapped atoms: $N=10$ (solid line), $N=10^3$ (dashed line), and
$N=10^6$ (dash-dotted line).}
\label{fig:Fig.2}
\end{figure}

\begin{figure}[ht]
\centerline{
\includegraphics[width=10cm]{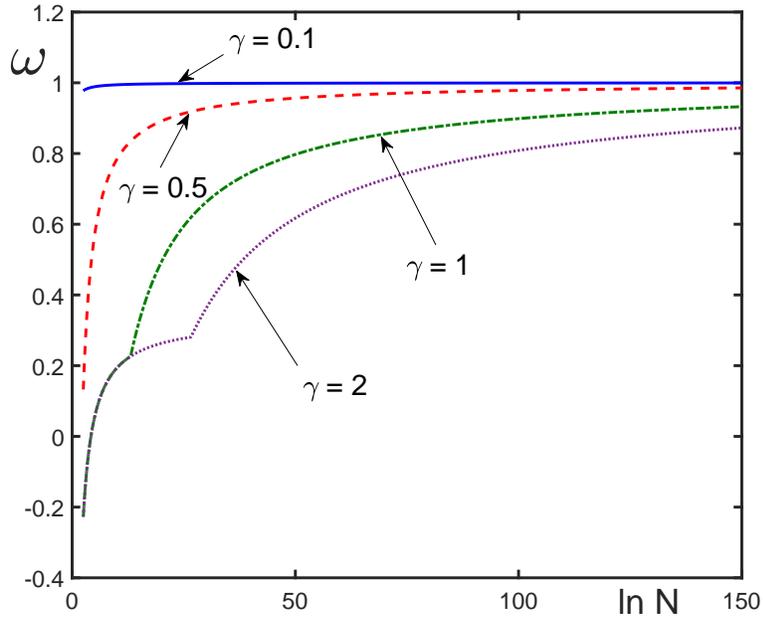} }
\vskip 3mm
\caption{Order index $\om(\hat{\rho}_1)$ as a function of $\ln N$ for different 
values of the gas parameter: $\gm=0.1$ (solid line), $\gm=0.5$ (dashed line), 
$\gm=1$ (dash-dotted line), and $\gm=2$ (dotted line).}
\label{fig:Fig.3}
\end{figure}

\end{document}